# Privacy protection of occupant behavior data and using blockchain for securely transferring temperature records in HVAC systems

Jingming Li[a], Nianping Li[a, *], Jinqing Peng[a], Zhibin Wu[a], Haijiao Cui[a]
[a]College of Civil Engineering, Hunan University, Changsha 410082, China

**Abstract**
The proportion of Energy consumption in the building industry is great, as well as the amount of cooling and heating system. Scholars have been working on energy conservation of Heating, ventilation, and air-conditioning and other systems in buildings. The application of occupant behavior data for building energy optimization has started gaining attention from scholars. However, occupant behavior data concerns many aspects of occupants' privacy. Different types of occupant behavior data contain occupants' private information to different levels. It is crucial to conduct privacy protection of occupant behavior data when using occupant behavior for energy conservation. This paper presents the aspects of privacy issue when using occupant behavior data, and methods to protect data privacy with blockchain technology. Both two options of using blockchain for privacy protection, sending data records as transactions and storing files on the blockchain, are explained and evaluated with temperature records from an open access paper. Sending data as transactions can be used between sensors and local building management system. While storing files on blockchain can be used for collaboration of different building management systems. Advantages, drawbacks, and potentials of using blockchain for data and file transfer are discussed. The results should be helpful for using occupant behavior data for building energy optimization.

# 1. Introduction

Buildings and buildings construction together are responsible for 36% of final energy consumption and around 40% of direct and indirect carbon emissions globally [1]. Heating, ventilation, and air-conditioning (HVAC) systems are the major cause of energy consumption in buildings. Air conditioning and space heating took up to 32% in residential end-use electricity consumption in America [2]. Scholars have been applying various methods to reduce energy consumption in HVAC. Since HVAC equipment is used to meet human comfort needs, studying occupant behavior becomes the beginning for the understanding complex in building energy consumption. As occupant behavior data starting being widely used in building energy conservation, the protection of occupant behavior data should be noticed.

## 1.1. Occupant behavior data

Before using occupant behavior data for building energy optimization, strategies were used for energy systems (i.e. district energy or renewable energy), control systems (i.e. lighting, shading control or HVAC control systems), building envelope (i.e. material or design of windows, walls, and other components) and simulations about these systems [3–8]. Liu et al. integrated demand response services to the smart grid, this method helped optimize HVAC loads and thermal storage [9]. Sembroiz et al. [10] presented an optimization for sensors and gateways in smart buildings. Hong et al. went further in studying occupant behavior by discussing occupant behavior in buildings for energy optimization. Hong et al. [11] classified energy-related occupant behavior as adaptive behaviors and non-adaptive behaviors based on factors of energy consumption and comfort. Both adaptive and non-adaptive behaviors are mainly human's response to thermal change, which ends to changes in HVAC loads. Peng et al. used machine learning for prediction of occupant behavior, and their results reached 7-52% energy saving [12].

Their works indicated that the use of the smart building in energy conservation had been promising, as well as the value of occupant behavior data. Hong et al. illustrated occupant behavior data (i.e. smart metering and smart building data, indoor and outdoor environmental data, occupant's interaction with the control system, occupancy data, survey, questionnaires, and self-reported data) as objective measurements and subjective measurements [11]. Both objective measurements and subjective measurements have privacy protection in their original fields. The problem is to securely manage collaboration of them after used for occupant behavior analysis.

## 1.2. Information privacy

As studies applying data and model approaches to understand occupant behavior, there is a lack of studies on privacy protection for occupant behavior data. While information privacy or data protection has been an issue in privacy concerns. Information privacy is about collecting and broadcasting data, technology, public expectations of privacy, legal and political issues around them [13–16]. Many studies have been working on information privacy. Wen et al. [17] scheduled workflows for cloud data privacy protection constraints. Wei et al. [18] presented automatic privacy protection for social images. Jiang et al. [19] compared performances of different Location Privacy Protection methods in Wireless Sensor Networks. The security of occupant behavior data has not been taken into enough considerations.

As mentioned earlier, Hong et al. listed occupant behavior data as five categories. Most data of the five categories could be used to identify, contact, or locate a single person, or to identify an individual in context [13,20–22]. Smart metering and smart building data could be used to track an occupant's activities. Building information modeling and geographic information system were integrated for visually monitoring occupant's activities [23,24]. Indoor and outdoor environmental data have been widely used energy-related building simulations. The environmental data can also be used for assisting trajectory tracking. Scholars in Princeton University used weather reports and four other types of information for location tracking [25]. Occupant's interaction with the control system could indicate occupant's location and daily schedules. Jia et al. used HVAC controllers as an example to present a tractable framework, using mutual information to balance data utilization and privacy protection [26]. Occupancy data [27], obtained by occupancy sensors, can be used for location and schedule tracking. The survey, questionnaires, and self-reported data [28] contain private information from respondents, including names, genders, living styles, and etc. for inference occupant behavior. Thus, privacy protection must be applied for occupant behavior data.

### 1.3. Data protection

HVAC systems contain lots of occupant behavior information, including indoor and outdoor environmental information, occupancy data, and occupant's interactions with controllers. Also, HVAC systems can cooperate with sensors for personalized ventilation and operation optimization. There have been studies about privacy protection in many areas. One of the most popular methods is to add random noise to original data. However, calculating noise is difficult and might result in information loss [26]. Mutual information is introduced to balance utility and privacy [29]. The method [26] extended mutual information and modeled occupancy and locations traces in Factorial Hidden Markov Model. The location traces were hidden behind aggregate occupancy information. Although occupant behavior can still be deducted from the distortion scheme. The objection of this paper is to provide a scheme for HVAC data and protecting occupant's privacy. Blockchain came into public in 2008 with Bitcoin [30]. For ten years, the Bitcoin network has been securely maintained by an open public blockchain. Projects have been carried for developing this technology for various applications [31]. Zyskind et al. [32] proposed a scheme which combines blockchain and off-blockchain data storage to protect privacy. Their method has proved the feasibility of using blockchain for privacy protection and the possibility of blockchain data processing.

Considering the importance of HVAC system data and technology advantages, this study presented a blockchain data scheme for HVAV data to utilize data and ensure privacy. There are basic foundations before choosing what kind of blockchain should be used [33]. In this case, a hybrid blockchain is needed since HVAC data for occupant behavior study requires shared access, The following contents explained this scheme. Section 2 explained the proposed solution of using blockchain for privacy protection. Section 3 evaluated the proposed method. The limits and future works are in Section 4. Section 5 concludes the paper.

### 2. Proposed Solution

The commonly used method is to store the hashed results on the blockchain. Using the raw data as input and a certain hash function, a hash is generated. The only thing stored on the blockchain is the hash of the data. Another method is to use sensors as wallets to send data as transactions.

## 2.1. Storing data as transactions

The transactions in digital coin usually seen as three parts, a transaction input, a transaction output, and an amount. The transaction model in digital coins mainly has two types, Unspent Transaction Output (UTXO) and Account Balance Model. Bitcoin transactions may have multiple transaction inputs and outputs for the adoption UTXO.

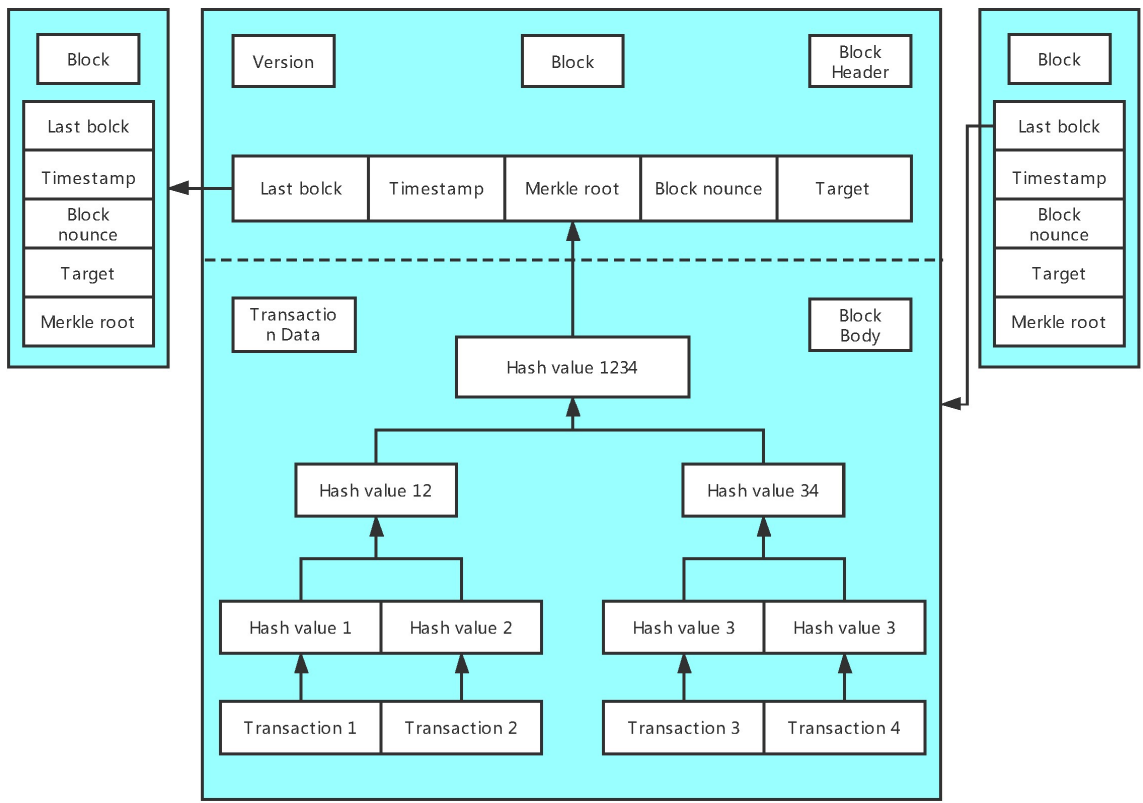

**Fig. 1.** The structure of a blockchain, based on the article by Greg Walker [33].

The overall available balance in a Bitcoin wallet is the sum of unspent outputs in the addresses owned by the user. Each time a transaction starts, the input is generated from prior transactions. Two outputs are generated when the transaction ends, one is the real amount coins, which go to the receiver's address and another is the change input, which goes back to the sender's wallet as an unspent transaction. By creating new addresses for each transaction, UTXO provides privacy for users. While another transaction model, Account Balance Model, which is used by Ethereum, resembles the current bank card. It tracks how much money each address has, and when spending money, the wallet checks its record to make sure that the sending address has enough balance before approving the transaction. When calculating balances, Account Balance Model can give an instant result without summing all addresses as UTXO.

Either using UTXO or Account Balance Model, the transactions are packed into blocks. Fig. 1 shows the structure of a Bitcoin block. The block structure of Ethereum resembles and contains other information, including the beneficiary, logs bloom, extra data and etc. All transactions are accessible in the blockchain with the sender and receiver encrypted. Therefore, by installing wallets on sensors and local Building Management System (BMS) as shown in Fig. 2, the transmission of indoor environment data to local management centers can be sent as transactions.

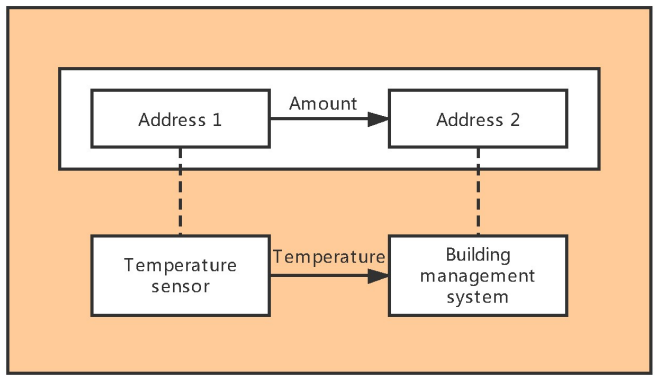

**Fig. 2.** Sending data as transactions on the blockchain.

In this study, the blockchains based on Account Balance Model are used, mainly for two reasons. Firstly, it is more efficient than UTXO. The simplicity in the Account Balance Model makes it easier to validate balance and transactions. Another reason is that Ethereum is suitable for developing complex smart contracts. The smart contracts can keep tracking records, and perform different tasks and collaborations based on them.

### 2.2. Storing files on the blockchain

Storing files on blockchain usually refers to storing hashes of data rather than storing data directly because the later one requires larger expenses to maintain a database [34]. In this study, this step is conducted using the InterPlanetary File System (IPFS), which is a distributed file system for devices over the Internet. The IPFS is a collective of IPFS objects. The structure of an IPFS object has two fields, a blob of unstructured binary data and an array of Link structures [35]. Links have three fields, the name of the link, the hash of the linked object, and the size of the linked IPFS object, including following its links; The data field is usually smaller than 256 kB with the list array being empty and there are header and footer in the content, otherwise, it is represented by a list of links and the names of the links would be empty [34]. When a file is uploaded to IPFS, it generates a unique hash value, which can be used to retrieve the file. Files larger than 256 kB will be separated to files smaller than 256 kB, the hashes of each small file will be generated as one hash value. This hash becomes the connection among the IPFS blocks as shown in Fig. 3.

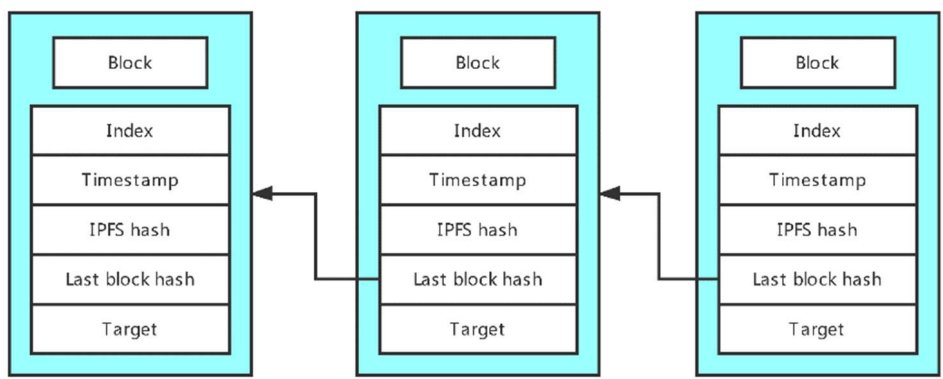

**Fig. 3.** The structure of IPFS blockchain, from the article by Coral Health [36].

However, IPFS seeks to connect all computing devices. Any person has the hash can access the file over the IPFS network. Thus, the file needs to be encrypted before uploading using GNU Privacy Guard (GPG). GPG has symmetric and asymmetric ciphers for encrypting and signing data and communications. Symmetric ciphers are mainly used to encrypt data. Asymmetric ciphers are mainly used to encrypt symmetric ciphers and digital signatures. In this study, symmetric ciphers are used to demonstrate the method. The file containing temperature records will be encrypted with the receiver's public key and then uploaded to IPFS. IPFS will create a unique hash for the file. Receivers can access the file with the hash and decrypt the file with the private key. The workflow is shown in Fig. 4.

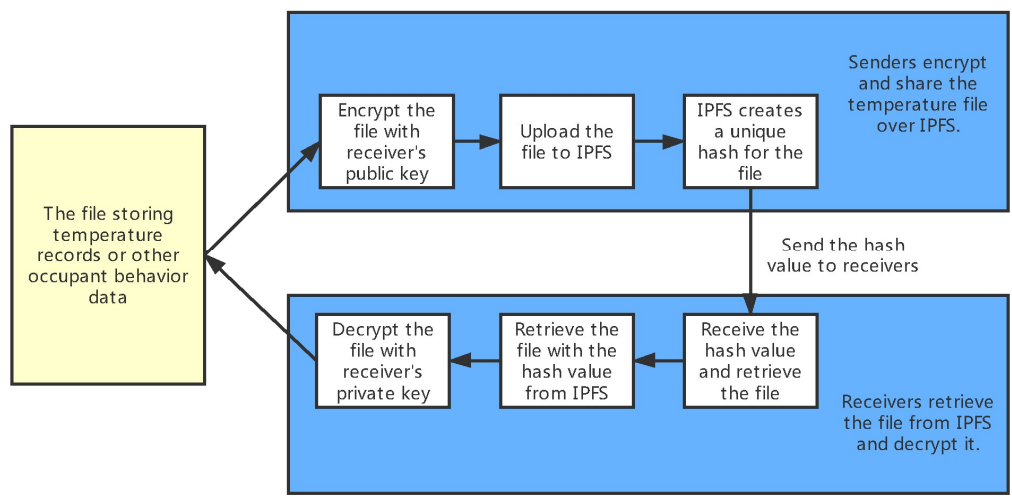

**Fig. 4.** The flow of sharing occupant temperature file on IPFS blockchain, based on the article by Coral Health [36].

More importantly, the directory structure of IPFS has the ability to deduplicate files automatically [35]. When modifications are made to the files, the previous directory has been replaced and the hash is regenerated.

### 3. Evaluation

To evaluate the methods, the data of sensor one in a university building from an open access paper was used [37] as the temperature records. A private Ethereum chain was created for evaluating sending temperature records as transactions. For the evaluation of storing files on the blockchain, both a cloud server and a local computer were used.

### 3.1. Transactions of temperature records

To conduct a transaction, two accounts were created. The account starting from '0xcc8e' was assigned as the sender and the account starting from '0xbb1d' was assigned as the receiver. Setting temperature as the amount for the transaction, the commands can be written as:

eth.sendTransaction ({from: eth.coinbase, to: eth.accounts[1], value: web3.toWei(22.9, "ether"), gas: 100000, gasPrice: 0})

A unique hash will be generated for this transaction. For the above transaction, it is:

0xa9da0346a437a0607a1c441fa26a84585b4f99572ef3044c990832e8d85fac4c.

All transactions can be checked from a blockchain explorer [38]. Fig. 5 shows the temperature records represented in transactions. Although the sender and receiver were replaced by hash values, the addresses in Fig. 5 were relatively lack of complexity. The limitations are discussed in Section 4.

| Tx Hash | Block | From | To | Value |
| --- | --- | --- | --- | --- |
| 0xa9da0346a437a0607a1c441fa26a84585b4f99572ef3044c990832e8d85fac4c | 715 | 0xcc8e730f18458694f5df7f1c296bdc4c65dcaa4e | 0xbd1d91ef6574a9d66f0ce8ca645447ae1c1d26fd | 22.9 |
| 0x5c88d6d9461a5c8f968292223eec4fabe4f33ba9514d9c0d6f4700e134aea8fb | 715 | 0xcc8e730f18458694f5df7f1c296bdc4c65dcaa4e | 0xbd1d91ef6574a9d66f0ce8ca645447ae1c1d26fd | 22.8 |
| 0x0f729e8a278a3e488d73287d4efafb77b9b2af096ff66c89165d6c1dcef766af | 715 | 0xcc8e730f18458694f5df7f1c296bdc4c65dcaa4e | 0xbd1d91ef6574a9d66f0ce8ca645447ae1c1d26fd | 22.7 |
| 0xec539d002f5968f9233fc1bbb2f27e736c5461cb3c26e82e766ded38fb8f515f | 715 | 0xcc8e730f18458694f5df7f1c296bdc4c65dcaa4e | 0xbd1d91ef6574a9d66f0ce8ca645447ae1c1d26fd | 22.6 |
| 0xc1cd09f3152c9ea4dde7514f8b89ff219a31c7e2d7f4da0963a584033aaef5dc | 715 | 0xcc8e730f18458694f5df7f1c296bdc4c65dcaa4e | 0xbd1d91ef6574a9d66f0ce8ca645447ae1c1d26fd | 22.6 |
| 0x5ca94981d9bc41bcfd9777ba99f533e1d1e04d42f1b428e24caab440de5d8f31 | 715 | 0xcc8e730f18458694f5df7f1c296bdc4c65dcaa4e | 0xbd1d91ef6574a9d66f0ce8ca645447ae1c1d26fd | 22.6 |
| 0x50257ba876fbba08972fa0849de40160e6c80a271c8d04e876c416d611d89a6b | 715 | 0xcc8e730f18458694f5df7f1c296bdc4c65dcaa4e | 0xbd1d91ef6574a9d66f0ce8ca645447ae1c1d26fd | 22.6 |
| 0xee48e52b5275eab75278af08c17f06e3cae196e990919f695be97e1c8dbb602f | 715 | 0xcc8e730f18458694f5df7f1c296bdc4c65dcaa4e | 0xbd1d91ef6574a9d66f0ce8ca645447ae1c1d26fd | 22.6 |
| 0x35e6f88ac32ad9a8bfc46352c50c3a53e91f4da4f7593b69d60d69df7f354938 | 715 | 0xcc8e730f18458694f5df7f1c296bdc4c65dcaa4e | 0xbd1d91ef6574a9d66f0ce8ca645447ae1c1d26fd | 22.8 |
| 0x5aefa01d6b3c3164886308ca4c4f9e9d0be3e37bbd79326b4efb0311b3d6d2a3 | 715 | 0xcc8e730f18458694f5df7f1c296bdc4c65dcaa4e | 0xbd1d91ef6574a9d66f0ce8ca645447ae1c1d26fd | 23.2 |
| 0xe48754793128fc4e25f17f0351c2fa8160e8c672070e88604897916301371734 | 715 | 0xcc8e730f18458694f5df7f1c296bdc4c65dcaa4e | 0xbd1d91ef6574a9d66f0ce8ca645447ae1c1d26fd | 23.5 |
| 0xc2eb7cb01c8e53edf6e47015c4230cd5c80c4a1c8fd0969a8c05bfda2d1f660c | 715 | 0xcc8e730f18458694f5df7f1c296bdc4c65dcaa4e | 0xbd1d91ef6574a9d66f0ce8ca645447ae1c1d26fd | 24.1 |
| 0xaef4eac6662bc9d759bdfc331a411aa14ac501daa6ec6871248455eca4eff3e5 | 715 | 0xcc8e730f18458694f5df7f1c296bdc4c65dcaa4e | 0xbd1d91ef6574a9d66f0ce8ca645447ae1c1d26fd | 24 |
| 0xcd7c3e19acff529034da968d299e1c04dc1b6d6ae704c425be5fbfaed2ae9a72 | 715 | 0xcc8e730f18458694f5df7f1c296bdc4c65dcaa4e | 0xbd1d91ef6574a9d66f0ce8ca645447ae1c1d26fd | 24.3 |
| 0x00e032626e2860ca0fa1e4ebc244aa3637be29dae70e449fea1febd6e995686dd | 715 | 0xcc8e730f18458694f5df7f1c296bdc4c65dcaa4e | 0xbd1d91ef6574a9d66f0ce8ca645447ae1c1d26fd | 23.8 |
| 0xc5322ae08fe364b79c6172286541b72c6ddcf94216722b0693ee4042c7e89f73 | 715 | 0xcc8e730f18458694f5df7f1c296bdc4c65dcaa4e | 0xbd1d91ef6574a9d66f0ce8ca645447ae1c1d26fd | 23.8 |
| 0xcfabd4c1ccb6724ed9c88e054faf061a4113354d7d8a2fa6f0329790e068b9f1 | 715 | 0xcc8e730f18458694f5df7f1c296bdc4c65dcaa4e | 0xbd1d91ef6574a9d66f0ce8ca645447ae1c1d26fd | 23.6 |
| 0xeeb544ae74d6b3887e700766a46ee157cdd2863f2dacc426cc576bc866cb3583 | 715 | 0xcc8e730f18458694f5df7f1c296bdc4c65dcaa4e | 0xbd1d91ef6574a9d66f0ce8ca645447ae1c1d26fd | 23.8 |
| 0x3d0fcd0e74d7be100492a50fb2a337cfe33b39bad8465cf52a14490bde9c1097 | 715 | 0xcc8e730f18458694f5df7f1c296bdc4c65dcaa4e | 0xbd1d91ef6574a9d66f0ce8ca645447ae1c1d26fd | 24 |
| 0xd7f4928e659cc76faded8300ead74248e1f7ecf24d52a0bf25e489aa89e71920 | 715 | 0xcc8e730f18458694f5df7f1c296bdc4c65dcaa4e | 0xbd1d91ef6574a9d66f0ce8ca645447ae1c1d26fd | 23.8 |
| 0xb2d9bcee1dc9524339fc0558d1451a4610623819a1a6a75fb8556d16691fd61a | 715 | 0xcc8e730f18458694f5df7f1c296bdc4c65dcaa4e | 0xbd1d91ef6574a9d66f0ce8ca645447ae1c1d26fd | 23.4 |
| 0x1fa8c7e9c51653fee31d85d5850631800d4d56a79c45e811b3f5a253ef0bc5cb | 715 | 0xcc8e730f18458694f5df7f1c296bdc4c65dcaa4e | 0xbd1d91ef6574a9d66f0ce8ca645447ae1c1d26fd | 23.3 |
| 0xe5709fe9ece0b0267bb1fe0c29fe2a51023beca99cf543fdd40817844fbd3767 | 715 | 0xcc8e730f18458694f5df7f1c296bdc4c65dcaa4e | 0xbd1d91ef6574a9d66f0ce8ca645447ae1c1d26fd | 23.2 |
| 0x9df091654348fae6b9b7868d85d4361b078f08b8bcb2f6ae1cc89b9b6e61728e | 715 | 0xcc8e730f18458694f5df7f1c296bdc4c65dcaa4e | 0xbd1d91ef6574a9d66f0ce8ca645447ae1c1d26fd | 23.1 |

**Fig. 5.** Transactions of temperature records.

### 3.2. Blockchain sortation of files containing temperature records

The other option of storing temperature record files in blockchain is also evaluated. The records were firstly encrypted and uploaded to IPFS network form the server named izwz9ccx74i196dnw7jldtz. The file is downloaded form IPFS on a Windows computer with the unique hash. Fig. 6 shows the process.

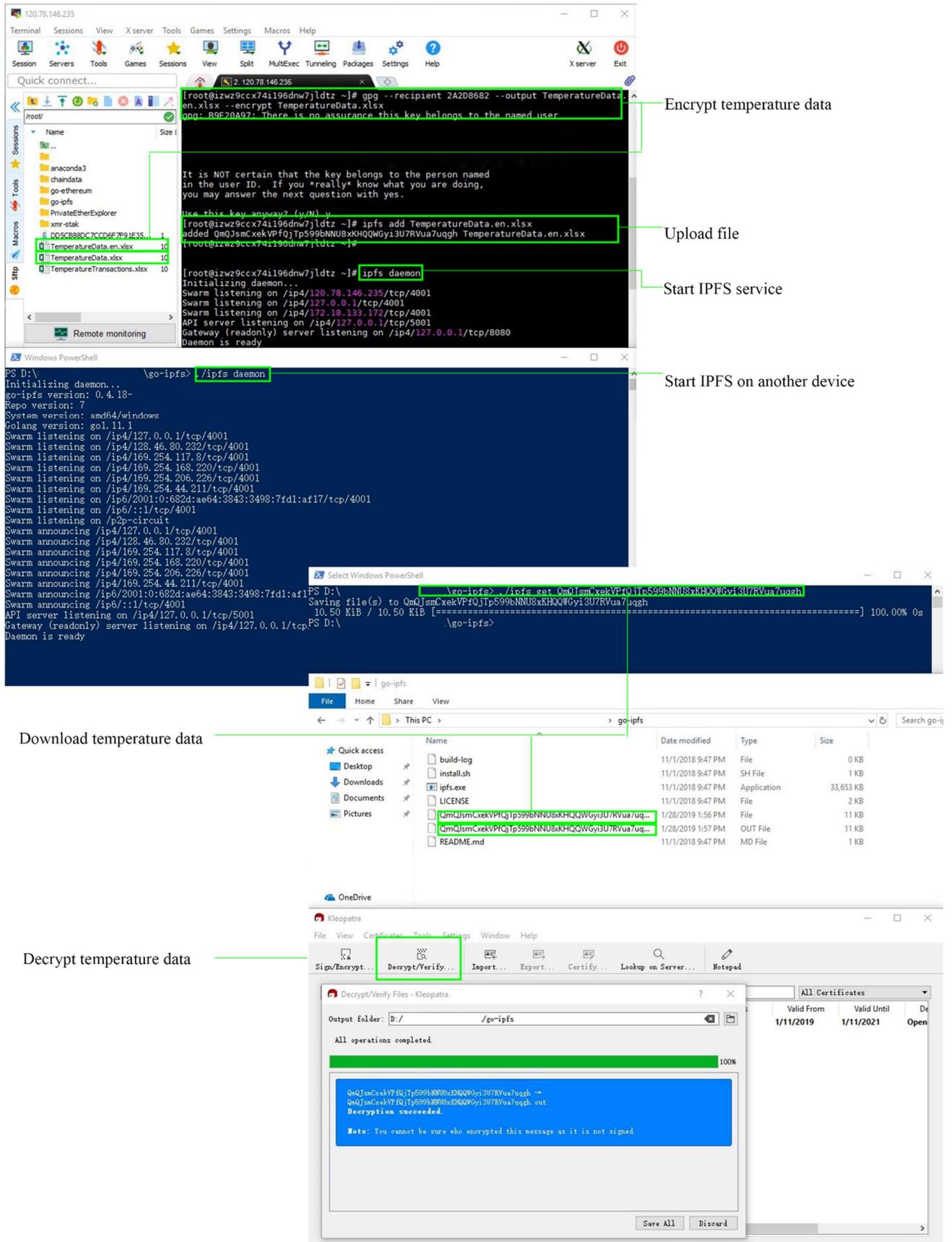

**Fig. 6.** Retrieving encrypted file from IPFS chain.

The only way to get this file is through this hash, and the key to decrypt the file is receiver's private key. This double protection ensures that files can only be accessed by designated people. In this case, building owners can strictly determine who to share their occupant behavior data with.

## 4. Discussions

As mentioned in the introduction, occupant behavior involves greatly in building energy consumption. Analysis of occupant data can benefit building energy optimization. It is certain that the protection of occupant data needs to be conducted. Two methods were proposed for the protection. Based on the above evaluations, both two options have advantage and drawbacks.

Using blockchain transactions to represent data can hide information of sensors behind hash. However, this method requires new addresses after a few transactions. The results in Fig. 5 indicated that the unchangeable address would be deductible. As for blockchain sortation of files containing occupant behavior data, the unique hash and encryption provide double protection for privacy, however, compared to data transfer, its operation is more suitable for file management.

### 4.1. Limitations

It is needed to note that this study has several limitations. The first is the simplicity in account addresses assigned to sensors and local computers. However, this problem can be solved by increasing the anonymity of the accounts. The addresses of senders and receivers can be automatically generated, or the addresses can be looped or rerouted through smart contracts. The second problem is IPFS is not an ideal tool for data collection, such as temperature records. The function of content sharing has not been fully utilized. Thirdly, this study only experimented on temperature records, many other data in the HVAC system should be tested in further studies. For the limit of source data, further experiments of securely energy optimization using smart contracts are not conducted.

### 4.2. Future directions

Although the two options have their drawbacks, they are the preliminary ideas for using blockchain for privacy protection of occupant behavior data. There are several topics needed to be studied, listed but not limited as followings. The first is to upgrade the sensors and local BMS computers so that its real-time data can be read and spread in the private chain. The second is to utilize the features of Ethereum and IPFS. Using smart contracts in Ethereum to enhance anonymity and conduct system optimization needs to be studied. Also, it is needed to study the integration of Ethereum private chain and IPFS, where the connection between sensors and local BMS computers are using the private chain and the BMS computers are communicating in IPFS. Thirdly, more data types should be tested, for instance, the signals of lighting control, security system, and others. When provided with sufficient data, energy optimization with privacy protection can be simulated.

## 5. Conclusions

This paper discussed using blockchain for privacy protection of occupant behavior data. This study has the following contributions:

1. This study discussed the privacy issue of occupant behavior data. Occupant behavior data concerns many areas in building management. It requires more studies when using occupant behavior for building energy optimization.

2. The reasons for choosing temperature records in the HVAC system has been explained and methods of using blockchain for privacy protection of occupant behavior data is discussed. Two options, both transferring data as transactions in blockchain and sharing data files through IPFS, have been evaluated.

3. Transferring data as transactions in blockchain requires more variations in the accounts to enhance anonymity. Sharing data files through IPFS needs more work on data transfer between sensors and local BMS, for instance, integrating private chain.

The limitations and possible future directions are also discussed in this paper. Adopting smart contracts for managing occupant behavior data might contribute to the protection of occupant privacy. The results should be helpful to properly use occupant data for building energy optimization.

**Acknowledgment**

This paper is supported by the China National Key R&D Program 'Solutions to heating and cooling of buildings in the Yangtze river region' (Grant No. 2016YFC0700305) and the Natural Science Foundation of China (No. 51878255). The authors are very grateful to the Natural Science Foundation of China and Hunan University Research Foundation for their support. J. Li would like to express his gratitude to the China Scholarship Council for the support (No. 201806130106) and everyone in Blockchain Network Theory of Purdue University.

**References**

[1] International Energy Agency, Buildings, (2017). https://www.iea.org/buildings/ (accessed November 3, 2018).

[2] EIA RECS, Residential Energy Consumption Survey (RECS) - Data - U.S. Energy Information Administration (EIA), (2017). doi:10.1109/PERCOMW.2004.1276898.

[3] M.S.M. Norhasri, M.S. Hamidah, A.M. Fadzil, Applications of using nano material in concrete: A review, Constr. Build. Mater. (2017). doi:10.1016/j.conbuildmat.2016.12.005.

[4] A.A.F. Husain, W.Z.W. Hasan, S. Shafie, M.N. Hamidon, S.S. Pandey, A review of transparent solar photovoltaic technologies, Renew. Sustain. Energy Rev. (2018). doi:10.1016/j.rser.2018.06.031.

[5] V.S.K.V. Harish, A. Kumar, A review on modeling and simulation of building energy systems, Renew. Sustain. Energy Rev. (2016). doi:10.1016/j.rser.2015.12.040.

[6] A. Tzempelikos, A.K. Athienitis, The impact of shading design and control on building cooling and lighting demand, Sol. Energy. (2007). doi:10.1016/j.solener.2006.06.015.

[7] A. Kirimtat, B.K. Koyunbaba, I. Chatzikonstantinou, S. Sariyildiz, Review of simulation modeling for shading devices in buildings, Renew. Sustain. Energy Rev. (2016). doi:10.1016/j.rser.2015.08.020.

[8] A. Lake, B. Rezaie, S. Beyerlein, Review of district heating and cooling systems for a

sustainable future, Renew. Sustain. Energy Rev. 67 (2017) 417–425. doi:10.1016/j.rser.2016.09.061.

[9] Y. Liu, N. Yu, W. Wang, X. Guan, Z. Xu, B. Dong, T. Liu, Coordinating the operations of smart buildings in smart grids, Appl. Energy. 228 (2018) 2510–2525. doi:10.1016/j.apenergy.2018.07.089.

[10] D. Sembroiz, D. Careglio, S. Ricciardi, U. Fiore, Planning and operational energy optimization solutions for smart buildings, Inf. Sci. (Ny). 0 (2018) 1–14. doi:10.1016/j.ins.2018.06.003.

[11] T. Hong, D. Yan, S. D'Oca, C.-F. Chen, Ten questions concerning occupant behavior in buildings: The big picture, Build. Environ. (2017). doi:10.1016/j.buildenv.2016.12.006.

[12] Y. Peng, A. Rysanek, Z. Nagy, A. Schlüter, Using machine learning techniques for occupancy-prediction-based cooling control in office buildings, Appl. Energy. 211 (2018) 1343–1358. doi:10.1016/j.apenergy.2017.12.002.

[13] D. DEPARTMENT OF VETERANS AFFAIRS Washington, Management of Data Breaches Involving Sensitive Personal Information (SPI), (2012). https://web.archive.org/web/20180710182048/http://web.archive.org/web/20150526030226/www.va.gov/vapubs/viewpublication.asp?pub_id=608 (accessed November 1, 2018).

[14] K. Michael, L. Johnston, Uberveillance and the Social Implications of Microchip Implants : Emerging Technologies, in: Uberveillance Soc. Implic. Microchip Implant. Emerg. Technol., 2014. doi:10.4018/978-1-4666-4582-0.

[15] Sensitive Information, (1996). https://www.its.bldrdoc.gov/fs-1037/dir-032/_4768.htm (accessed November 1, 2018).

[16] Wikipedia, Information privacy, (n.d.). https://en.wikipedia.org/wiki/Information_privacy#cite_note-2 (accessed November 2, 2018).

[17] Y. Wen, J. Liu, W. Dou, X. Xu, B. Cao, J. Chen, Scheduling workflows with privacy protection constraints for big data applications on cloud, Futur. Gener. Comput. Syst. (2018). doi:10.1016/j.future.2018.03.028.

[18] Z. Wei, Y. Wu, Y. Yang, Z. Yan, Q. Pei, Y. Xie, J. Weng, AutoPrivacy: Automatic privacy protection and tagging suggestion for mobile social photo, Comput. Secur. 76 (2018) 341–353. doi:10.1016/j.cose.2017.12.002.

[19] J. Jiang, G. Han, H. Wang, M. Guizani, A survey on location privacy protection in Wireless Sensor Networks, J. Netw. Comput. Appl. (2018). doi:10.1016/j.jnca.2018.10.008.

[20] Wikipedia, Personally identifiable information, (n.d.). https://en.wikipedia.org/wiki/Personally_identifiable_information (accessed November 2, 2018).

[21] G. Stevens, CRS Report for Congress Data Security Breach Notification Laws, (2012).

https://fas.org/sgp/crs/misc/R42475.pdf.

[22] S.S. Greene, Security program and policies : principles and practices, 2014.

[23] M. Afkhamiaghda, M. Mahdaviparsa, K. Afsari, T.L. McCuen, Occupants Behavior-Based Design Study Using BIM-GIS Integration: An Alternative Design Approach for Architects: Proceedings of the 35th CIB W78 2018 Conference: IT in Design, Construction, and Management, in: I. MUTIS, T. Hartmann (Eds.), Adv. Informatics Comput. Civ. Constr. Eng., Springer International Publishing, 2019. doi:10.1007/978-3-030-00220-6.

[24] Y. Song, X. Wang, Y. Tan, P. Wu, M. Sutrisna, J. Cheng, K. Hampson, Trends and Opportunities of BIM-GIS Integration in the Architecture, Engineering and Construction Industry: A Review from a Spatio-Temporal Statistical Perspective, ISPRS Int. J. Geo-Information. 6 (2017) 397. doi:10.3390/ijgi6120397.

[25] A. Mosenia, X. Dai, P. Mittal, N.K. Jha, PinMe: Tracking a smartphone user around the world, IEEE Trans. Multi-Scale Comput. Syst. 4 (2018) 420–435. doi:10.1109/TMSCS.2017.2751462.

[26] R. Jia, R. Dong, S.S. Sastry, C.J. Spanos, Privacy-enhanced architecture for occupancy-based HVAC Control, Proc. 8th Int. Conf. Cyber-Physical Syst. - ICCPS '17. (2017) 177–186. doi:10.1145/3055004.3055007.

[27] T. Hong, S.C. Taylor-Lange, S. D'Oca, D. Yan, S.P. Corgnati, Advances in research and applications of energy-related occupant behavior in buildings, Energy Build. (2016). doi:10.1016/j.enbuild.2015.11.052.

[28] T.A. Nguyen, M. Aiello, Energy intelligent buildings based on user activity: A survey, Energy Build. (2013). doi:10.1016/j.enbuild.2012.09.005.

[29] F. Du Pin Calmon, N. Fawaz, Privacy against statistical inference, 2012 50th Annu. Allert. Conf. Commun. Control. Comput. Allert. 2012. (2012) 1401–1408. doi:10.1109/Allerton.2012.6483382.

[30] S. Nakamoto, Bitcoin: A Peer-to-Peer Electronic Cash System, (2008) 9. doi:10.1007/s10838-008-9062-0.

[31] J. Evans, Bitcoin 2.0: Sidechains And Ethereum And Zerocash, Oh My!, (2014). https://techcrunch.com/2014/10/25/bitcoin-2-0-sidechains-and-zerocash-and-ethereum-oh-my/ (accessed November 4, 2018).

[32] G. Zyskind, O. Nathan, A.S. Pentland, Decentralizing privacy: Using blockchain to protect personal data, Proc. - 2015 IEEE Secur. Priv. Work. SPW 2015. (2015) 180–184. doi:10.1109/SPW.2015.27.

[33] B. Suichies, Why Blockchain must die in 2016, Medium. (2018). https://medium.com/@bsuichies/why-blockchain-must-die-in-2016-e992774c03b4 (accessed November 13, 2018).

[34] ConsenSys, An Introduction to IPFS – ConsenSys – Medium, (n.d.). https://medium.com/@ConsenSys/an-introduction-to-ipfs-9bba4860abd0 (accessed

January 4, 2019).

[35] J. Benet, IPFS - Content Addressed, Versioned, P2P File System, (2014). http://arxiv.org/abs/1407.3561 (accessed January 7, 2019).

[36] Coral Health, Learn to securely share files on the blockchain with IPFS!, (n.d.). https://medium.com/@mycoralhealth/learn-to-securely-share-files-on-the-blockchain-with-ipfs-219ee47df54c (accessed December 28, 2018).

[37] O. Irulegi, A. Serra, R. Hernández, Data on records of indoor temperature and relative humidity in a University building, Data Br. 13 (2017) 248–252. doi:10.1016/J.DIB.2017.05.029.

[38] Blockchain Explorer for Ethereum Private blockchain, (n.d.). https://github.com/adeshshukla/PrivateEtherExplorer (accessed December 29, 2018).